\documentstyle[11pt,aaspp4]{article}

\begin{document}

\title{Identification of Galactic Wind Candidates using Excitation Maps: \\
Tunable-Filter Discovery of a Shock-Excited Wind in the Galaxy NGC~1482}

\author{Sylvain Veilleux\altaffilmark{1} and David S. Rupke}

\affil{Department of Astronomy, University of Maryland, College Park,
MD 20742; veilleux,drupke@astro.umd.edu}

\altaffiltext{1}{Cottrell Scholar of the Research Corporation}

\begin{abstract}
Multi-line imaging of the nearby disk galaxy NGC~1482 with the Taurus
Tunable Filter (TTF) on the Anglo-Australian Telescope reveals a
remarkable hourglass-shaped [N~II] $\lambda$6583/H$\alpha$ excitation
structure suggestive of a galactic wind extending at least 1.5 kpc
above and below the disk of the host galaxy. Long-slit spectroscopy
confirms the existence of a large-scale outflow in this galaxy. The
entrained wind material has [N~II] $\lambda$6583/H$\alpha$ ratios in
excess of unity while the disk material is characterized by H~II
region-like line ratios indicative of a starburst.  Expansion
velocities of order 250 km s$^{-1}$ are detected in the wind material,
and a lower limit of 2 $\times$ 10$^{53}$ ergs is derived for the
kinetic energy of the outflow based on the gas kinematics and the
amount of ionized material entrained in the outflow. This is the first
time to our knowledge that a galactic wind is discovered using
excitation maps. This line ratio technique represents a promising new
way to identify wind galaxy candidates before undergoing more
time-consuming spectroscopic follow-ups. This method of selection may
be particularly useful for samples of galaxies at moderate redshifts.
\end{abstract}


\section{Introduction}

In an effort to better constrain the morphology, kinematics, and
origin of the warm ionized gas on the outskirts of galaxies we have
obtained deep emission-line images of several nearby starburst and
active galaxies using the Taurus Tunable Filter (TTF) on the 3.9m
Anglo-Australian and 4.2m William Herschel Telescopes (see Veilleux
2001 for more detail).  In the course of this study, we discovered a
remarkable emission-line structure in the early-type spiral galaxy
NGC~1482. This galaxy has so far received relatively little attention
in the literature. It is classified as a peculiar SA0/a in the Revised
3rd Catalogue (De Vaucouleurs et al. 1991) based primarily on the
presence of a dust lane across the disk of the galaxy. Located at a
distance of 19.6 Mpc (Tully 1988), it is an infrared-bright galaxy
with log[L$_{\rm IR}$/L$_\odot$] = 10.5 (e.g., Soifer et al. 1989;
Sanders, Scoville, \& Soifer 1991), which is rich in molecular gas and
dust (e.g., Sanders et al. 1991; Young et al. 1995; Chini et al.
1996) and is undergoing vigorous star formation (e.g., Moshir et al.
1990; Devereux \& Hameed 1997; Thornley et al. 2000).  In a recent
emission-line imaging survey of early-type spirals, Hameed \& Devereux
(1999) noticed the presence in NGC~1482 of ``filaments and/or chimneys
of ionized gas extending perpendicular to the disk.''  The present
study expands on the results of Hameed \& Devereux, using deeper
emission-line maps at H$\alpha$ and [N~II] $\lambda$6583 and
complementary long-slit spectra. We find that the [N~II]/H$\alpha$
excitation map of this galaxy is particularly useful at distinguishing
between the star-forming disk and the entrained, shock-excited wind
material.  This excitation signature could be used in the future to
more efficiently identify powerful galactic winds in the local and
distant universe.

\section{Observations}

NGC 1482 was observed on the night of 2000 December 16 using the
``blue'' Taurus Tunable Filter (TTF; Bland-Hawthorn \& Jones 1998) at
the Anglo-Australian Telescope (AAT). This instrument was used in the
``charge shuffling / band switching'' mode to maximize
sensitivity to faint flux levels.  The basic idea is to move charge up
and down within the detector at the same time as switching between two
discrete wavelengths with the tunable filter.  In this way, the
on-band image is obtained nearly simultaneously as the off-band
(continuum) image.  For our observations, the charges were moved every
minute and the chip was read out after a total integration time of 32
minutes (i.e. 16 minutes on-band and 16 minutes off-band). Four sets
of observations were obtained of NGC~1482: two centered on redshifted
H$\alpha$ for a total on-band integration time of 32 minutes, and two
centered on redshifted [N~II] $\lambda$6583 for the same duration. The
continuum images for each set of observations were obtained in a
straddle mode, where the off-band image is made up of a pair of images
that ``straddle'' the on-band image in wavelength; this greatly
improves the accuracy of the continuum removal since it corrects for
slopes in the continuum and underlying absorption features (see also
Maloney \& Bland-Hawthorn 2001).  The wavelength calibration was
checked before and after each pair of observations and was found to be
stable. The bandpass of the TTF was set to 14.6 \AA\ throughout the
observations. This bandpass was chosen to separate H$\alpha$
$\lambda$6563 emission from [N~II] $\lambda\lambda$6548, 6583
emission, while still producing a monochromatic field of view of
several arcminutes, much larger than the total extent of
NGC~1482. Phase effects due to the angular Fabry-Perot interferometer
response can therefore be safely ignored in the rest of the
discussion. The MITLL2A CCD was used for these observations, providing
a pixel scale of 0$\farcs$37 pixel$^{-1}$ and a field of view of
10$\arcmin$ $\times$ 10$\arcmin$. The observations were obtained under
photometric and $\sim$ 1$\farcs$5 seeing conditions.

On 2001 September 19 -- 21, several complementary long-slit spectra
were obtained of NGC~1482. The Dual-Beam Spectrograph (DBS) on the
MSSSO 2.3m telescope was used at the Nasmyth f/17.9 focus with the
1200B and 1200R gratings. The CCD was a SITe 1752 $\times$ 532
$\times$ 15$\mu$ pixel device. At a plate scale 0\farcs 91 pix$^{-1}$,
a 2\arcsec\ slit gave a resolution of 1.2 \AA\ FWHM and a wavelength
coverage of 4400 -- 5400 \AA\ and 6100 -- 7050 \AA. On one occasion,
the slit was aligned along the minor axis of the galaxy
(P.A. $\approx$ 13$^\circ$; De Vaucouleurs et al. 1991) passing
through the nucleus.  In all other cases, the slit was positioned along
103$^\circ$ and offset from 0 to $\pm$ 16\arcsec\ from the major axis
of the galaxy.  The exposure times were typically 1200 seconds.

\section{Results}

Figure 1 shows the distribution of the H$\alpha$ and [N~II]
$\lambda$6583 emission in NGC~1482. Strong H$\alpha$ and [N~II]
emission is detected along the plane of the host galaxy
(P.A. $\approx$ 103$\arcdeg$). In addition, an hourglass-shaped
structure is seen in both H$\alpha$ and [N~II] $\lambda$6583,
extending along the minor axis of the galaxy at least $\sim$ 1.5 kpc
above and below the galactic plane. This structure is more easily
visible in [N~II] $\lambda$6583 than in H$\alpha$. This is
particularly apparent in the lower left panel of Figure 1, where we
present a [N~II]/H$\alpha$ ratio map of this object. The [N~II]
$\lambda$6583/H$\alpha$ ratios measured in the disk of the galaxy
[$\approx$ 0.3 (outer disk) -- 0.6 (inner disk)] are typical of
photoionization by stars in H~II regions, but the ratios in the
hourglass structure are 3 -- 7 times larger ([N~II]
$\lambda$6583/H$\alpha$ $\approx$ 1.0 -- 2.3). This ratio of a
collisionally excited line to a recombination line is fundamentally a
measure of the relative importance of heating and ionization
(e.g., Osterbrock 1989).

[N~II]/H$\alpha$ ratios enhanced relative to H~II regions are often
observed in the extraplanar material of normal disk galaxies,
including our own (e.g., Rand, Kulkarni, \& Hester 1990; Veilleux et
al. 1995a; Reynolds, Haffner, \& Tufte 1999; Hoopes, Walterbos, \&
Rand 1999; Rossa \& Dettmar 2000; Collins \& Rand 2001; Miller 2002).
Although photoionization by the hardened and diluted radiation field
from OB stars in the disk is partially responsible for these peculiar
line ratios (e.g., Sokolowski 1992; Bland-Hawthorn, Freeman, \& Quinn
1997), a secondary source of ionization or heating is often needed to
explain in detail the runs of line ratios in these galaxies (e.g.,
Reynolds et al. 1999; Collins \& Rand 2001; Miller 2002). The extreme
[N~II] $\lambda$6583/H$\alpha$ ratios in the hourglass structure of
NGC~1482 also require an additional source of heating.  Ruling out
photoionization by an AGN based on the lack of evidence for genuine
nuclear activity in NGC~1482 (e.g., Kewley et al. 2000), the most
likely explanation for these unusual line ratios is shock ionization.
Interaction of an energetic large-scale outflow with the ambient
material of a gas-rich host galaxy will cause shock waves with
velocities of 100 -- 500 km s$^{-1}$.  The shocks will produce a
strong flux of EUV and soft X-ray radiation which may be absorbed in
the shock precursor H~II region (e.g., Dopita \& Sutherland
1995). Arguably one of the best examples known of shock-excited
nebulae associated with large-scale galactic winds is the kpc-scale
superbubble in the galaxy NGC~3079 (e.g., Filippenko \& Sargent 1992;
Veilleux et al. 1994; Cecil et al. 2001). Large [N~II]
$\lambda$6583/H$\alpha$ ratios are observed throughout the bubble,
reaching values of $\sim$ 3 at the base of the bubble where the widths
of the emission lines exceeds 400 km s$^{-1}$. H~II region-like line
ratios are observed everywhere else in the disk (Veilleux et
al. 1995a).

The kinematics of the line-emitting gas derived from our long-slit
spectroscopy of NGC~1482 confirm that the hourglass structure is due
to a large-scale galactic wind and that the extreme line ratios are
produced through shocks. Figure 2 shows the disturbed kinematics in
the extraplanar material. Line splitting of up to $\sim$ 250 km
s$^{-1}$ is detected along the axis of the hourglass structure out to
at least 16\arcsec\ (1.5 kpc) above and below the galaxy disk. Normal
galactic rotation dominates the kinematics of the gas within 5 --
6\arcsec\ ($\sim$ 500 pc) from the disk. Maximum line splitting often
coincides with regions of low emission-line surface brightness. These
results can be explained if the extraplanar emission-line material
forms a biconical edge-brightened structure which is undergoing
outward motion away from the central disk. In this case, the
blueshifted (redshifted) emission-line component corresponds to the
front (back) surface of the bicone. The lack of obvious velocity
gradient in the centroid of the line emission suggests that the main
axis of the bicone lies close to the plane of the sky. The fact that
the amplitude of the line splitting does not decrease significantly with
distance from the galaxy indicates that the entrained material is not
experiencing significant deceleration (i.e., it is a blown-out wind). 

The mass involved in this wind can be estimated from the total
H$\alpha$ emission outside of the disk. Defining the wind material as
having [N~II] $\lambda$6583/H$\alpha$ $\ge$ 1, we get a wind H$\alpha$
flux (luminosity) of $\sim$ 1.7 $\times$ 10$^{-13}$ erg s$^{-1}$
cm$^{-2}$ ($\sim$ 8.0 $\times$ 10$^{39}$ erg s$^{-1}$) with an
uncertainty of about $\pm$ 25 -- 30\%. The [S~II]
$\lambda$6731/$\lambda$6716 line ratios measured from our long-slit
spectra indicate that the entrained material has a density $n_e \la
100$ cm$^{-3}$. The mass in entrained line-emitting material is
therefore $\ga$ 3.6 $\times$ 10$^5$ $n_{e,2}^{-1}$ M$_\odot$ where
$n_{e,2}$ is normalized to 100 cm$^{-3}$ (assuming Case B
recombination and an effective recombination coefficient for H$\alpha$
of 8.6 $\times$ 10$^{-14}$ cm$^{-3}$ s$^{-1}$; Osterbrock 1989). A
correction for possible dust extinction intrinsic to NGC~1482 will
further increase this number (note that the extinction due to our
Galaxy, E(B--V) = 0.04, is negligible; Schlegel et al. 1998).  A
simple kinematic model can be used to estimate a few key wind
parameters.  The bisymmetric morphology and kinematics of the wind
nebula in NGC~1482 suggest that the entrained material lies on the
surface of a cylindrically symmetric bicone.  If most of the gas
motion is tangential to the surface of the bicone and given that the
axis of the bicone lies approximately in the plane of the sky, the
radial outflow velocity $V_{\rm out}$ = $\Delta~V$ / (2 $\cdot$
sin~$\theta$), where $\Delta~V$ is the observed maximum line splitting
and $\theta$ is the angle between the surface and the main axis of the
bicone.  Taking $\Delta~V$ $\approx$ 250 km s$^{-1}$ and $\theta$
$\approx$ 30$^\circ$ based on the morphology of the wind (Fig. 1), we
get $V_{\rm out}$ $\approx$ 250 km s$^{-1}$. Given that $\Delta~V$ and
$\theta$ are observed to be approximately constant throughout the wind
nebula, the total kinetic energy involved in the outflow is $\ga$ 2
$\times$ 10$^{53}$ $n_{e,2}^{-1}$ ergs.

The dynamical timescale of the outflow can be estimated from the
outflow velocity and radial extent: $\tau_{\rm dyn} \approx R / V_{\rm
out} \approx 6 \times 10^6$ years.  Combining the dynamical timescale
and kinetic energy of the outflow, we derive a time-averaged kinetic
energy injection rate of $\ga$ 1 $\times$ 10$^{39}$ $n_{e,2}^{-1}$
ergs s$^{-1}$.  This can be compared with the energy injection rate
from star formation in NGC~1482 of $\sim$ 7 $\times$ 10$^{42}$ $f_{\rm
out}$~L$_{\rm IR, 11}$ $\approx$ 2 $\times$ 10$^{42}$ $f_{\rm out}$
erg s$^{-1}$, where L$_{\rm IR,11}$ is the infrared luminosity in
units of 10$^{11}$ L$_\odot$ and $f_{\rm out}$ is the fraction of the
star-forming disk which is contributing to the outflow (Veilleux et
al. 1994, eqn. 12).  The broad base of the wind nebula near the galaxy
disk (Fig. 1) suggests that the source of energy for the wind is
distributed over a region of $\sim$ 2 kpc, i.e. $f_{\rm out}$ $\approx$
50\% of the total extent of the star-forming (H$\alpha$-emitting)
portion of the disk.  The starburst at the base of the wind nebula in
NGC~1482 is therefore powerful enough to drive the outflow as long as
$n_e$~$\ga$~0.1~cm$^{-3}$.

\section{On the Use of Excitation Maps to Identify Starburst-Driven Wind
Galaxies}

The traditional method of identifying galaxy-scale winds in starburst
galaxies is to look for the kinematic signature (e.g., line splitting)
of the wind along the minor axis of the host galaxy disk. Edge-on disk
orientation reduces contamination of the wind emission by the
underlying disk material and facilitates the identification. This
method is time-consuming since it requires deep spectroscopy of each
candidate wind galaxy with spectral resolution of $\la$ 100 km
s$^{-1}$. Line ratio maps like the one shown in Figure 1 represent a
promising new way to detect galactic winds in starburst galaxies. The
line ratio method only requires taking narrow-band images of candidate
wind galaxies centered on two (or more) key diagnostic emission lines
which emphasize the contrast in the excitation properties between the
shocked wind material and the star-forming disk of the host galaxy.
[N~II] $\lambda$6583/H$\alpha$, [S~II] $\lambda\lambda$6716,
6731/H$\alpha$, and [O~I] $\lambda$6300/H$\alpha$ are the optical line
ratios of choice for $z \la 0.5$ galaxies (these ratios are enhanced
in the wind of NGC~1482), while [O~II] $\lambda$3727/H$\beta$ and
[O~II] $\lambda$3727/[O~III] $\lambda$5007 could be used for objects
at larger redshifts.  The spatial resolution of these images must be
sufficient to distinguish the galaxy disk from the wind
material. Using NGC~1482 as a template, we find that
high-[N~II]/H$\alpha$ winds in edge-on starburst galaxies would still
be easily detected out to a distance of $\sim$ 200 Mpc under 1\arcsec\
resolution. Imagers equipped with adaptive optics systems should be
able to extend the range of these searches by an order of magnitude.

This method relies on the dominance of shock excitation in the optical
line-emitting wind component.  Surveys of local powerful wind galaxies
(e.g., Heckman, Armus, \& Miley 1990; Bland-Hawthorn 1995; Veilleux et
al. 1995b; Lehnert \& Heckman 1996; Veilleux 2001) confirm that shocks
generally are the dominant source of excitation in the wind
material. These shock-dominated wind nebulae present line ratios which
are markedly different from the star-forming disks of the host
galaxies.  The case of the superbubble in NGC~3079 has already been
discussed in \S 3. At the other end of the excitation spectrum is the
wind in M~82. The [N~II]/H$\alpha$ map of the southern wind lobe of
M~82 (Fig. 4 of Shopbell \& Bland-Hawthorn 1998) presents two distinct
fan-like structures with H~II region-like ratios originating from the
two bright star-forming regions in the disk of this galaxy. The line
ratio technique would not be able to distinguish between line emission
from this type of photoionization-dominated wind nebulae and
contamination from a star-forming disk seen nearly face-on.  Blind
searches for galactic winds based on the excitation contrast between
the disk and wind components would therefore favor the detection of
winds in edge-on hosts where the wind component is not projected onto
the disk component. This orientation bias would need to be taken into
account to get a complete census of starburst-driven wind galaxies.
AGN-driven winds may also contaminate samples selected from excitation
maps if the spatial resolution is not sufficient to separate the 
active nucleus from the disk material.

\acknowledgements
The authors wish to thank R. B. Tully who brought to our attention the
peculiar properties of NGC~1482.  We also thank J. Bland-Hawthorn for
help in using the Taurus Tunable Filter and for entertaining
discussions.  The authors acknowledge partial support of this research
by a Cottrell Scholarship awarded by the Research Corporation,
NASA/LTSA grant NAG 56547, and NSF/CAREER grant AST-9874973.

\clearpage

\begin{figure*}[ht]
\caption{ Narrow-band images of NGC~1482 obtained with the TTF:
(clockwise from upper left corner) red continuum, [N~II] $\lambda$6583
line emission, H$\alpha$ line emission, and [N~II]
$\lambda$6583/H$\alpha$ ratio.  The panels on the right are on a
logarithmic intensity scale, while those on the left are on a linear
scale.  North is at the top and east to the left.  The positions of
the continuum peaks are indicated in each image by two crosses. The
spatial scale, indicated by a horizontal bar at the bottom of the red
continuum image, is the same for each image and corresponds to $\sim$
21\arcsec, or 2 kpc for the adopted distance of 19.6 Mpc for NGC~1482.
The [N~II] $\lambda$6583/H$\alpha$ ratio is below unity in the galaxy
disk but larger than unity in the hourglass-shaped nebula above and
below the disk. This structure is highly suggestive of a galactic
wind.}
\end{figure*}

\begin{figure*}[ht]
\caption{ Sky-subtracted long-slit spectra obtained parallel to the
galactic disk (P.A. $\sim$ 103\arcdeg). For each panel, south-east is
to the left and north-west to the right. The spectra displayed on the
left are offset to the north-east by 0\arcsec\ (bottom panel),
9\arcsec\ -- 10\arcsec, 11\arcsec\ -- 12\arcsec, and 13 -- 14\arcsec\
(top panel) from the major axis of the host galaxy. The spectra
displayed on the right are offset by approximately the same quantity
in the south-west direction. The vertical segment in each panel
represents 500 km s$^{-1}$. The presence of line splitting above and
below the disk confirms the presence of a large-scale wind in this
galaxy. }
\end{figure*}

\clearpage

\end{document}